\documentclass[aps,prl,preprint,preprintnumbers,superscriptaddress]{revtex4-1}
\usepackage{bm}% bold math
\usepackage{lipsum}% http://ctan.org/pkg/lipsum
\usepackage{graphicx,amssymb}
\usepackage{dcolumn}% Align table columns on decimal point
\usepackage{bm}% bold math

\usepackage{color}
\usepackage{ulem}
\usepackage[colorlinks=true,linkcolor=blue,citecolor=blue,urlcolor=blue]{hyperref}
\usepackage{tabularx}
\usepackage{multirow}
\usepackage[utf8] {inputenc}

\begin{document}
%\draft \preprint{SLAC-draft}

\title{Coincident onset of charge density wave order at a quantum critical point in underdoped YBCO}

\author{H. Jang}
\affiliation{Stanford Synchrotron Radiation Lightsource, SLAC National Accelerator Laboratory, Menlo Park, CA 94025}
\author{W.-S. Lee}\email{leews@stanford.edu}
\affiliation{Stanford Institute for Materials and Energy Science, SLAC National Accelerator Laboratory and Stanford University, Menlo Park, CA 94025}
\author{S. Song}
\affiliation{Linac Coherent Light Source, SLAC National Accelerator Laboratory, Menlo Park, CA 94025}
\author{H. Nojiri}
\author{S. Matsuzawa}
\author{H. Yasumura}
\affiliation{Institute for Materials Research, Tohoku University, Katahira 2-1-1, Sendai, 980-8577, Japan}
\author{H. Huang}
\author{Y.-J. Liu}
\affiliation{Stanford Synchrotron Radiation Lightsource, SLAC National Accelerator Laboratory, Menlo Park, CA 94025}
\author{J. Porras}
\affiliation{Max-Planck-Institut f{\"u}r Festk{\"o}rperforschung, Heisenbergstrasse 1, D-70569 Stuttgart, Germany}
\author{M. Minola}
\affiliation{Max-Planck-Institut f{\"u}r Festk{\"o}rperforschung, Heisenbergstrasse 1, D-70569 Stuttgart, Germany}
\author{B. Keimer}
\affiliation{Max-Planck-Institut f{\"u}r Festk{\"o}rperforschung, Heisenbergstrasse 1, D-70569 Stuttgart, Germany}
\author{J. Hastings}
\author{D. Zhu}
\affiliation{Linac Coherent Light Source, SLAC National Accelerator Laboratory, Menlo Park, CA 94025}
\author{T. P. Devereaux}
\affiliation{Stanford Institute for Materials and Energy Science, SLAC National Accelerator Laboratory and Stanford University, Menlo Park, CA 94025}
\author{Z.-X. Shen}
\affiliation{Stanford Institute for Materials and Energy Science, SLAC National Accelerator Laboratory and Stanford University, Menlo Park, CA 94025}
\affiliation{Geballe Laboratory for Advanced Materials, Departments of Physics and Applied Physics, Stanford University, Stanford, CA 94305}
\author{C.-C. Kao}
\affiliation{SLAC National Accelerator Laboratory, Menlo Park, CA 94025}
\author{J.-S. Lee}\email{jslee@slac.stanford.edu}
\affiliation{Stanford Synchrotron Radiation Lightsource, SLAC National Accelerator Laboratory, Menlo Park, CA 94025}

\date{\today}

\begin{abstract}
The recently demonstrated x-ray scattering approach using a free electron laser with a high field pulsed magnet has opened new opportunities to explore the charge density wave (CDW) order in cuprate high temperature superconductors. Using this approach, we substantially degrade the superconductivity with magnetic fields up to 33 T to investigate the onset of CDW order in YBa$_2$Cu$_3$O$_x$ at low temperatures near a putative quantum critical point (QCP) at $p_1\sim $ 0.08 holes per Cu. We find no CDW can be detected in a sample with a doping concentration less than $p_1$. Our results indicate that the onset of the CDW ground state lies inside the zero-field superconducting dome, and broken translational symmetry is associated with the putative QCP at $p_1$.
\end{abstract}

%\pacs{78.70.Ck; Need more}

% 78.70.Ck	X-ray scattering

\maketitle

%%%%%%%%%%%%%%%%%%%%%%%%%%%%%%%%%%%%%%%%%%%%%%

The existence of multiple electronic ordering phenomena in cuprate high temperature superconductors has stimulated the need to identify and characterize the associated quantum critical points (QCP).  In the archetypal cuprate, YBa$_2$Cu$_3$O$_x$  (YBCO), thermodynamic and transport measurements have hinted at the existence of additional QCPs inside the superconductivity (SC) dome\cite{LeBoeuf2013, LeBoeuf2011,LeBoeuf2013,Wu2011,Yu2016,Sebastian2010,Ramshaw2015,Grissonnanche2014}. For example, the sign change of the Hall coefficient at low temperatures, which implies a Fermi surface reconstruction, only occurs for doping concentrations $p > p_1 \sim 0.08$ holes/Cu \cite{LeBoeuf2011}. In addition, quantum oscillation measurements suggest that the electron effective mass of YBCO diverges around $p_1$ \cite{Sebastian2010, Ramshaw2015}. This QCP conjecture is further supported by resistivity measurements of YBCO under high magnetic fields ($H$) \cite{Grissonnanche2014}, where YBCO's SC regime at $H$ $=$ 30 T shrinks and separates into two domes [6], centered at the putative QCPs at $p  \sim p_1$ and $\sim$ 0.19, respectively. While the putative QCP at $p \sim$ 0.19 has received more attention due to its possible connection with the mysterious pseudogap state \cite{Badoux2016}, the QCP at $p_1$ is relatively less studied. If the anomalies at $p_1$ are indeed related to a QCP, it would be crucial to identify any associated broken symmetries.

The discovery of an incommensurate charge density wave (CDW) in YBCO \cite{Ghiringhelli2012,Chang2012, Hawthorn2012, Hayden2013,Keimer2014, Blanco2014,Gerber2015,Chang2016,Jang2016} and other hole-doped cuprate families \cite{Haceker2013,Croft2014,Comin2014,Neto2014, Tabis2014,Neto2015} has raised an intriguing question about its relevance for the anomalies at $p_1$ in YBCO. From x-ray scattering studies of YBCO at $H$ $=$ 0 \cite{Blanco2014}, the CDW is only detectable for $p$ $>$ $p_1$, consistent with the association of $p_1$ as the onset doping concentration of the CDW order. However, since the presence of SC at low temperatures disrupts the CDW order, the actual onset doping concentration of the CDW and its connections to those putative QCPs inferred from transport measurements are still obscure. To overcome this issue, it is necessary to destabilize superconducting long-range order at low temperatures by a high magnetic field. Notably, these conditions were achieved in a recent sound velocity measurement in high magnetic fields \cite{Lalibert2017}, suggesting that the CDW indeed onsets at $p_1$. Nevertheless, in order to firmly establish this notion, a direct measurement of CDW order using x-ray scattering at high magnetic field near $p_1$ is extremely desirable.

In this study, we address the relationship between the CDW and SC orders across a putative QCP at $p_1$ by using x-ray scattering under high magnetic fields capable of suppressing superconductivity in this low doping range. We find no trace of CDW at a doping concentration $p$ $\sim$ 0.07 (\textit{i.e.} less than $p_1$), in spite of the fact that SC long-ranger order is heavily degraded by the magnetic field. Upon increasing doping to $p$ $\sim$ 0.08, we find that the CDW becomes detectable at low temperatures with $H$ $=$ 33 T, but it is absent in zero magnetic field at all temperatures.  At $p$ $\sim$ 0.09 (\textit{i.e.} slightly higher than $p_1$), the CDW evolves into a three dimensionally ordered state at high magnetic fields, like those previously reported at higher doping \cite{Gerber2015,Chang2016,Jang2016}. These findings unambiguously establish that the CDW, which breaks translational symmetry in high magnetic fields, onsets at $p \sim$ $p_1$, \textit{i.e.} at the putative QCP.

High-quality, detwinned ortho-II YBa$_2$Cu$_3$O$_x$ single crystals with $x$ $=$ 6.40, 6.43, and 6.48 were selected for this work, corresponding to doping concentrations of $p$ $=$ 0.07 ($T_c$ = 35 K), 0.08 $\sim$ $p_1$ (43 K), and 0.09 (54 K), respectively. The experiment was carried out at the XCS instrument of the Linac Coherent Light Source \cite{Alonso2015}, to employ the setup that synchronizes the x-ray free electron laser pulses and pulsed magnetic fields \cite{Gerber2015,Jang2016}. Magnetic fields up to 33 T were used to suppress superconductivity to allow for the detection of CDW order at low temperatures. The magnetic field is applied along the $c$-axis (\textit{i.e.} perpendicular to the CuO$_2$ planes) with a lowest achievable sample temperature of 10 K. 

We first focus on the $x$ = 6.48 sample with a doping concentration $p$ = 0.09, which is slightly higher than $p_1$ = 0.08 and is also close to the lowest doping concentration at which the CDW diffraction pattern can be resolved by resonant soft x-ray scattering at the Cu $L_3$-edge at zero magnetic field \cite{Blanco2014}. Figure 1 (a) shows the zero-field data taken at $T$ $=$ $T_c$ (54 K) at which the CDW intensity should be maximal because the SC state has not formed to compete with CDW. Consistent with previous results \cite{Blanco2014}, we observe a rod-like diffraction pattern elongated along the $l$-direction in the reciprocal space as marked by the dashed-line box in Fig. 1(a), indicating that the CDW is poorly correlated along the \textit{c}-axis (\textit{i.e.} out-of-CuO$_2$ plane) and thus, is quasi-two-dimensional. To study the weak CDW peak, we average intensity profiles along the $k$-direction from $l$ $=$ 0.3 and 0.7 reciprocal lattice units (hereafter, r.l.u.), as shown in the lower panel of Fig. 1(a). The peak profile is centered around $Q$ = (0, 2-$q$, 1/2) with $q$ $\sim$ 0.33 r.l.u. Figure 1(b) shows the diffraction pattern after cooling the sample down to the superconducting state at 10 K. As expected, on the influence of SC, the CDW signal weakens and becomes undetectable at our signal-to-noise (i.e. detection) level. 

By applying magnetic fields up to 30 T to suppress SC in the $p$ $\sim$ 0.09 sample, the CDW intensity exhibits a significant  enhancement centered at $l$ $=$ 1, as shown in the upper panel of Fig. 1(c). In addition, the diffraction pattern is no longer rod-like as in zero field, but is much more concentrated in both $k$- and $l$- directions, indicating that the CDW develops longer range correlations not only within a CuO$_2$ plane, but also between planes (\textit{i.e.} along the $c$-axis). In other words, the CDW becomes three-dimensionally (3D) ordered. The field dependence of the average intensity near $l = 1$ is shown in the lower panel of Fig. 1(c). We observe that the 3D order starts to emerge at an onset field that lies in the range of 25 $\sim$ 30 T, while the CDW intensity at lower fields was primarily due to the enhancement of the quasi-2D order. The behavior of the quasi-2D and 3D CDW orders is consistent with previous x-ray scattering measurements in high fields on YBCO at higher doping concentrations \cite{Gerber2015,Chang2016,Jang2016}. As argued in previous studies \cite{Jang2016, Chang2016}, these two seemingly different CDWs share the same origin -- an unidirectional and long-range CDW correlation which can manifest either 2D or 3D character by tuning the magnetic field.

After confirming the existence of a CDW order just above $p_1$, we now investigate the CDW in the more underdoped regime at $p = 0.08 \sim p_1$. Figure 2(a) shows the averaged intensity from $l$ $=$ 0.4 to 0.6 r.l.u. taken at $T$ $=$ $T_c$ (43 K, upper panel) and 10 K (black markers in the lower panel) in zero magnetic field. Consistent with a previous resonant soft x-ray scattering result on YBCO samples grown in the same manner \cite{Blanco2014}, we could not resolve any signature of the CDW order in this range of reciprocal space at these two temperatures. In fact, even for magnetic fields up to 30 T, no CDW patterns can be unambiguously identified. However, at our maximum field, $H =$ 33 T, a weak trace of CDW diffraction peak at $Q$ $\sim$ (0, 2-$q$, 1/2) with a $q$ $\sim$ 0.3 can be resolved (red markers in the lower panel of Fig. 2(a)). As shown in the $kl$-plane intensity map and its averaged intensity [Fig. 2(b)], the primary order at 33 T is still quasi 2D with no sign of a 3D CDW at $l = 1$. Possibly, even higher magnetic fields are required for the 3D CDW to develop. Nevertheless, this finding demonstrates that the CDW instability still exists at $p = 0.08$, despite the fact the CDW is absent at all temperatures without applying a high magnetic field to suppress the SC.

Figure 3 (left-inset) shows high-field x-ray scattering data on our most underdoped sample, $p$ $=$ 0.07 ($T_c$ = 35 K). Obviously, no signature of the CDW is resolved at the lowest temperature (10 K) and in the maximum magnetic field (33 T). Since the sample is not superconducting at this temperature and field \cite{Grissonnanche2014}, CDW is unlikely to emerge at an even larger field. Furthermore, $H_{C2}$ of the $p$ = 0.07 compound is known to be lower than that of $p$ = 0.08 compound \cite{Grissonnanche2014}; thus, a higher field is actually required to reveal the CDW in the $p$ = 0.08 compound. The fact that the CDW signal can be resolved in $p$ = 0.08 compound but not in the $p$ = 0.07 indicates that the CDW correlation has disappeared at this doping concentration.

Finally, taking our data together, Figure 3 shows the integrated intensity of the CDW peak profile at 10 K and at $\sim$ 30 T as a function of doping. Note that the data points were obtained by averaging the region of reciprocal space containing both the quasi-2D and 3D CDW orders to reflect the amplitude of the order parameter. Apparently, the CDW amplitude, $\Delta_{CDW}$, weakens with decreasing doping. At $p$ $=$ 0.08, the CDW is absent at zero magnetic field, but emerges at low temperature when SC is heavily suppressed by the applied magnetic field. Eventually, the CDW instability vanishes for doping concentrations less than $p$ $=$ 0.07. This indicates a critical doping $p = p_{\rm CDW} \sim p_1$ for CDW formation, above which the CDW amplitude grows monotonically, indicative of a continuous quantum phase transition.   

To put our findings in context, we superimpose our data with other relevant measurements as shown in Fig. 4. First, the onset doping of the CDW ($p_{\rm CDW}$) in YBCO appears somewhat higher than that of SC ($p_{SC} \sim 0.05$). Although the difference in the critical $p$ may depend on the detection limit of the CDW peak in the x-ray diffraction pattern, these data suggest that the CDW emerges inside the SC dome. Second, the thermodynamic transition indicated by ultrasound ($T_V$)\cite{Lalibert2017} and Hall coefficient ($T_L$)\cite{LeBoeuf2011}, which has been attributed to the transition to the high field CDW state, extrapolates to zero as $p$ approaches $p_{\rm CDW}$, corroborating the notion that the CDW transition is continuous. Third, this integrated phase diagram also suggests an intriguing interplay between the CDW and spin density wave (SDW) in YBCO. Inelastic neutron scattering (INS) found the formation of an incommensurate SDW after the system loses the long range antiferromagnetic (AFM) order due to doping \cite{Haug2010}. The SDW transition temperature ($T_{SDW}$ in Fig. 4) decreases when approaching $p_{\rm CDW}$, indicating that SDW and CDW compete with each other. This is consistent with prior resonant x-ray scattering work in zero magnetic field on Zn-doped YBCO compounds \cite{Blanco2013}. This competition may arise from the fact that the ordering directions of the two orders are orthogonal. Their relationship and the difference to the stripe order found in La-based cuprates are further discussed in a recent theoretical study \cite{Nie2017}. Intriguingly, extrapolating the trend of $T_{\rm SDW}$, one would expect a QCP at $p_{\rm SDW}$ where the SDW disappears. However, since the $T_{SDW}$ is not sharply defined \cite{Haug2010}, and it is hard to quantify the uncertainty on doping concentration for samples from different experiments, it still remains unclear whether the $p_{\rm SDW}$ is larger than, or equal to $p_{\rm CDW}$. Nevertheless, our results lead to a sharp question regarding whether there are one or two QCPs in the proximity of $p_1$. Moreover, both $T_c$ at high field and the upper critical field for long-range SC order are maximal at $\sim p_1$ \cite{Grissonnanche2014}, indicating that quantum critical fluctuations of one or the other may boost superconductivity. The answer to the aforementioned question would identify the relevant quantum fluctuations and provide further insight into the superconducting pairing mechanism. 

%%%%%%%%%%%%%%%%%%%%%%%%%%%%%%%%%%%%%%%%%%%%%%
%ACKNOWLEDGMENTS
%%%%%%%%%%%%%%%%%%%%%%%%%%%%%%%%%%%%%%%%%%%%%%

We thank S. A. Kivelson for insightful discussions. This work was supported by the Department of Energy (DOE), Office of Science, Basic Energy Sciences, Materials Sciences and Engineering Division, under contract DE-AC02-76SF00515. X-ray FEL studies were carried out at the Linac Coherent Light Source, a Directorate of SLAC and an Office of Science User Facility operated for the U.S. DOE, Office of Science by Stanford University. For characterizing all samples, resonant soft x-ray scattering measurements were carried out at the Stanford Synchrotron Radiation Lightsource (BL13-3), a Directorate of SLAC and an Office of Science User Facility operated for the U.S. DOE, Office of Science by Stanford University. H.N. and S.M.  acknowledges the support by KAKENHI 23224009, 15K13510, ICC-IMR and MD-program.

%%%%%%%%%%%%%%%%%%%%%%%%%%%%%%%%%%%%%%%%%%%%%%

%%%%%%%%%%%%%%%%%%%%%%%%%%%%%%%%%%%%%%%%%%%%%%

%%%%%%%%%%%%%%%%%%%%%%%%%%%%%%%%%%%%%%%%%%%%%%
\begin{figure}
\begin{center}
\includegraphics[width=0.72\textwidth]{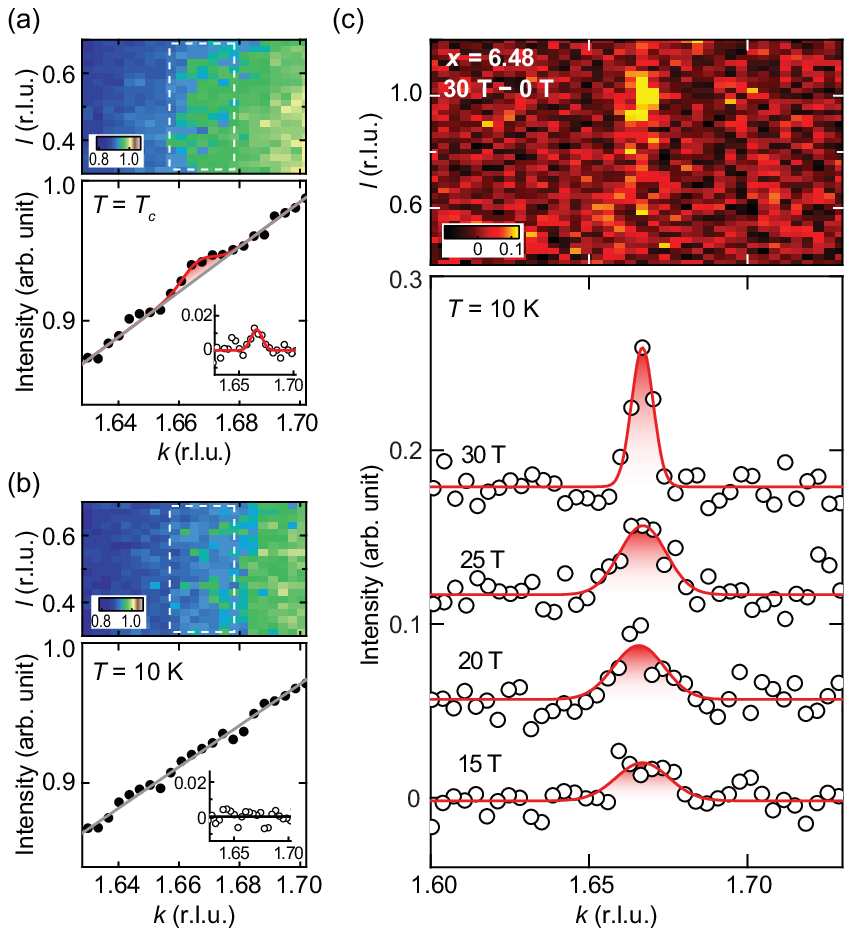}
\caption{(color online) CDW in YBa$_2$Cu$_3$O$_x$, $x$ = 6.48 ($p$ = 0.09). (a) The upper panel shows the scattering intensity map in the $kl$-plane at $h$ $=$ 0 taken at $T$ $=$ $T_c$ in the absence of any magnetic field. The white dashed box indicates the location for the quasi-2D CDW. The lower panel shows the averaged intensity from $l$ $=$ 0.3 to 0.7 r.l.u. The red curve is a Gaussian fit to the peak profile with a linear background (grey line). The inset show the background subtracted averaged intensity. (b) Same as (a), but the data were taken in the superconducting state at $T$ $=$ 10 K. The black line in the inset is a guide to the eye for zero intensity. (c) The upper panel shows the difference map in the $kl$-plane at $h$ = 0 obtained by subtracting the data taken at $H$ $=$ 0 T from the $H$ = 30 T data. The measurement temperature is 10 K. The lower panels show the averaged intensity of difference maps from $l$ $=$ 0.8 to 1.2 r.l.u. at representative magnetic fields, which are vertically shifted for visual clarity. The red curves are Gaussian fits to peak profiles.} \label{Fig1}
\end{center}
\end{figure}
%%%%%%%%%%%%%%%%%%%%%%%%%%%%%%%%%%%%%%%%%%%%%%

%%%%%%%%%%%%%%%%%%%%%%%%%%%%%%%%%%%%%%%%%%%%%%
\begin{figure}[t]
\includegraphics[width=0.70\textwidth]{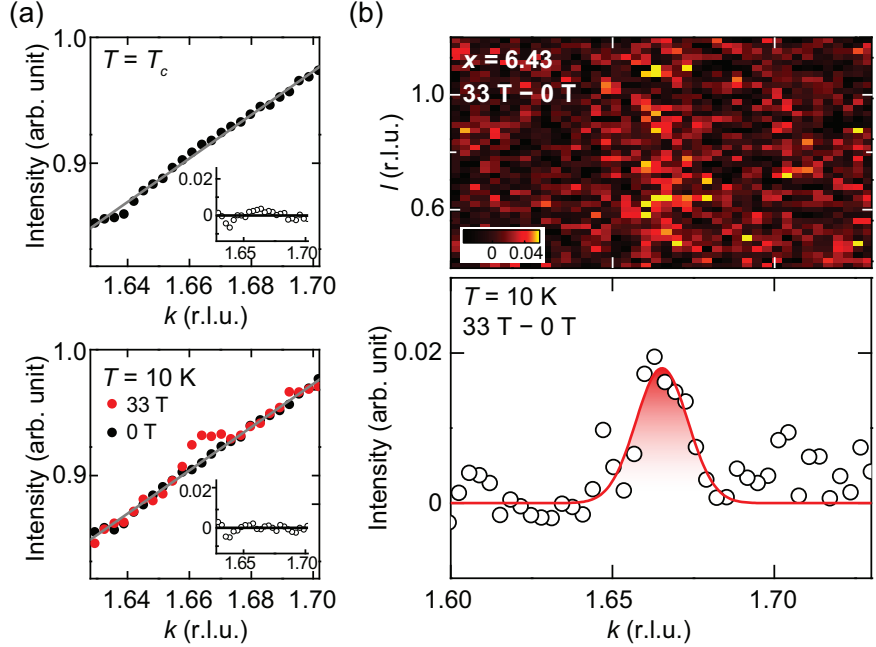}
\caption{(color online) CDW in YBa$_2$Cu$_3$O$_x$, $x$ = 6.43 ($p$ = 0.08). (a) Averaged intensity between $l$ $=$ 0.3 to 0.7 r.l.u. in the $kl$-plane at $h$ $=$ 0 taken at zero magnetic field at $T$ $=$ $T_c$ (upper) and at $T$ $=$ 10 K (lower, black markers). The averaged intensity taken at $T$ $=$ 10 K and $H$ $=$ 33 T is also plotted as the red markers in the lower panel. The intensity profiles after removing the background are shown in insets in which the black line serves as a guide-to-the-eye for zero intensity. (b) The upper panel shows the difference map in the $kl$-plane at $h$ $=$ 0 obtained by subtracting the data taken at $H$ = 0 T from the $H$ $=$ 33 T data. The measurement temperature is 10 K. The lower panel shows the averaged intensity of the difference map from $l$ $=$ 0.4 to 1.2 r.l.u.. The red curve is a Gaussian fit to the peak profile.}
\label{Fig2}
\end{figure}
%%%%%%%%%%%%%%%%%%%%%%%%%%%%%%%%%%%%%%%%%%%%%%

%%%%%%%%%%%%%%%%%%%%%%%%%%%%%%%%%%%%%%%%%%%%%%
\begin{figure}[b]
\includegraphics[width=0.78\textwidth]{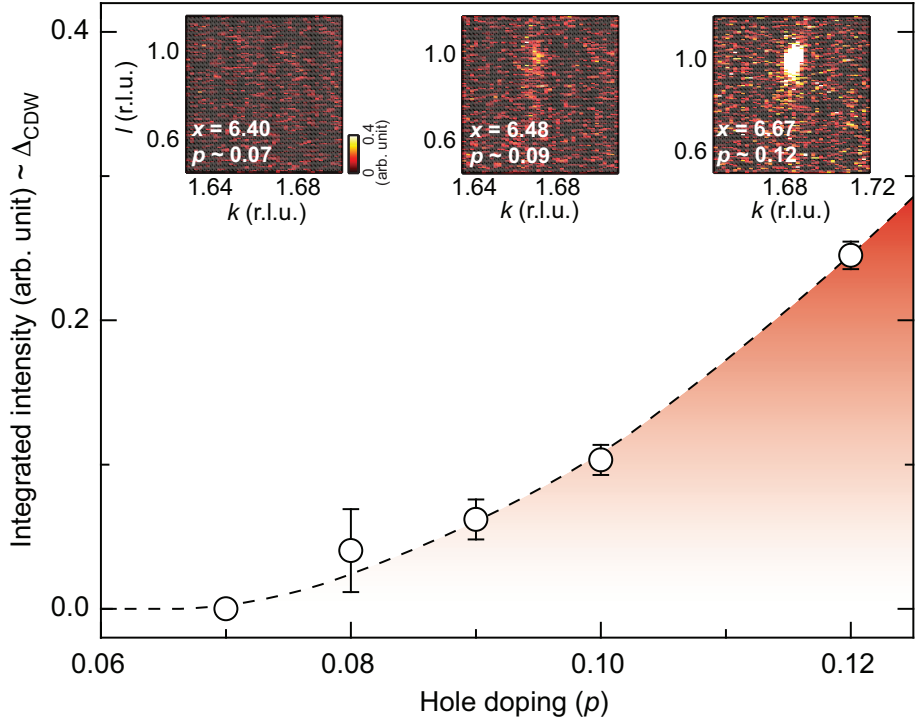}
\caption{(color online) Doping dependence of the integrated peak intensity that is obtained by computing the peak area in the averaged peak profile (as defined for the lower panel of Fig. 2(b)). The integrated peak intensity reflects the CDW amplitude $\Delta_{CDW}$. The black dashed curve is a guide-to-the-eye. Insets show difference maps in the $kl$-plane at $h$ = 0 for three representative doping concentrations. These maps were obtained by subtracting data taken at $H$ = 0 T from the $H$ $\sim$ 30 T data. The data for $p$ $\geq$ 0.1 were obtained from the same date set published in Ref. \cite{Jang2016}.}
\label{Fig3}
\end{figure}
%%%%%%%%%%%%%%%%%%%%%%%%%%%%%%%%%%%%%%%%%%%%%%

%%%%%%%%%%%%%%%%%%%%%%%%%%%%%%%%%%%%%%%%%%%%%%
\begin{figure}[t]
\includegraphics[width=0.72\textwidth]{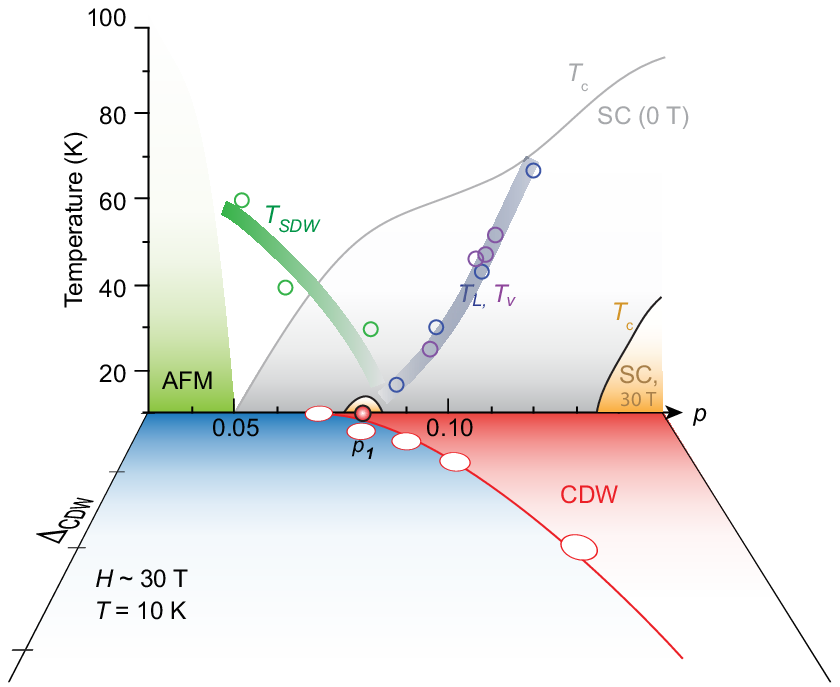}
\caption{(color online) A sketch of the YBCO phase diagram in the heavily underdoped regime with several other measurements superimposed. $T_c$ ($H$ $=$ 0) (grey curve) is the SC transition temperature at zero magnetic field. The orange shaded regions are the SC region at $H$ $=$ 30 T \cite{Grissonnanche2014}. $T_{\rm SDW}$ is the transition temperature of SDW measured by INS \cite{Haug2010}. $T_L$ is the temperature scale at which Hall coefficient changes to negative \cite{LeBoeuf2011}. $T_V$ is the onset temperature of the anomaly observed in sound velocity \cite{Lalibert2017}. The data shown in Fig. 3 is plotted in the orthogonal plane to the phase diagram.}
\label{Fig4}
\end{figure}
%%%%%%%%%%%%%%%%%%%%%%%%%%%%%%%%%%%%%%%%%%%%%%


\begin{thebibliography}{}

\bibitem{LeBoeuf2011} D. LeBoeuf, N. Doiron-Leyraud, B. Vignolle, M. Sutherland, B. J. Ramshaw, J. Levallois, R. Daou, F. Lalibert\'{e}, O. Cyr-Choini\`{e}re, J. Chang, Y. J. Jo, L. Balicas, R. Liang, D. A. Bonn, W. N. Hardy, C. Proust, and L. Taillefer, Phys. Rev. B \textbf{83}, 054506 (2011).

\bibitem{LeBoeuf2013} D. LeBoeuf, S. Kr\"{a}mer, W. N. Hardy, Ruixing Liang, D. A. Bonn, and Cyril Proust, Nature Physics \textbf{9}, 79 (2013).

\bibitem{Wu2011} T. Wu \textit{et al.}, Nature \textbf{477}, 191 (2011).

\bibitem{Yu2016} F. Yu, \textit{et al.}, PNAS \textbf{113}, 12667 (2016).
 
%\bibitem{Grissonnanche2015} G. Grissonnanche, F. Laliberte, S. Dufour-Beausejour, A. Riopel, S. Badoux, M. Caouette-Mansour, M. Matusiak, A. Juneau-Fecteau, P. Bourgeois-Hope, O. Cyr-Choiniere, J. C. Baglo, B. J. Ramshaw, R. Liang, D. A. Bonn, W. N. Hardy, S. Kramer, D. LeBoeuf, D. Graf, N. Doiron-Leyraud, L. Taillefer,  https://arxiv.org/abs/1508.05486

\bibitem{Sebastian2010} Suchitra E. Sebastian, N. Harrison, M. M. Altarawneh, C. H. Mielke, Ruixing Liang, D. A. Bonn, W. N. Hardy, and G. G. Lonzarich, Proc. Nat. Acad. Sci.  \textbf{107}, 6175 (2010).

\bibitem {Ramshaw2015} B. J. Ramshaw1, S. E. Sebastian, R. D. McDonald, James Day, B. S. Tan, Z. Zhu, J. B. Betts, Ruixing Liang, D. A. Bonn, W. N. Hardy, N. Harrison, Science \textbf{348} 317 (2015).


\bibitem{Grissonnanche2014} G. Grissonnanche, O. Cyr-Choini\`{e}re, F. Lalibert\'{e}, S. Ren\'{e} de Cotret, A. Juneau-Fecteau, S. Dufour-Beaus\'{e}jour, M.-\`{E}. Delage, D. LeBoeuf, J. Chang, B. J. Ramshaw, D. A. Bonn, W. N. Hardy, R. Liang, S. Adachi, N. E. Hussey, B. Vignolle, C. Proust, M. Sutherland, S. Kr\"{a}mer, J.-H. Park, D. Graf, N. Doiron-Leyraud, and L. Taillefer, Nat. Commun. \textbf{5}, 3280 (2014).

\bibitem{Badoux2016} S. Badoux, W. Tabis, F. Lalibert\'{e}, G. Grissonnanche, B. Vignolle, D. Vignolles, J. B\'{e}ard, D. A. Bonn, W. N. Hardy, R. Liang, N. Doiron-Leyraud, L. Taillefer, and C. Proust, Nature (London) \textbf{531}, 210 (2016).

\bibitem{Ghiringhelli2012} G. Ghiringhelli, M. Le Tacon, M. Minola, S. Blanco-Canosa, C. Mazzoli, N. B. Brookes, G. M. De Luca, A. Frano, D. G. Hawthorn, F. He, T. Loew, M. Moretti Sala, D. C. Peets, M. Salluzzo, E. Schierle, R. Sutarto, G. A. Sawatzky, E. Weschke, B. Keimer, and L. Braicovich, Science \textbf{337}, 821 (2012).

\bibitem{Chang2012} J. Chang, E. Blackburn, A. T. Holmes, N. B. Christensen, J. Larsen, J. Mesot, R. Liang, D. A. Bonn, W. N. Hardy, A. Watenphul, M. v. Zimmermann, E. M. Forgan, and S. M. Hayden, Nat. Phys. \textbf{8}, 871 (2012).

\bibitem{Hawthorn2012} A. J. Achkar, R. Sutarto, X. Mao, F. He, A. Frano, S. Blanco-Canosa, M. Le Tacon, G. Ghiringhelli, L. Braicovich, M. Minola, M. M. Sala, C. Mazzoli, R. Liang, D. A. Bonn, W. N. Hardy, B. Keimer, G. A. Sawatzky, and D. G. Hawthorn, Phys. Rev. Lett. \textbf{109}, 167001 (2012).

\bibitem{Hayden2013} E. Blackburn, J. Chang, M. Hücker, A. T. Holmes, N. B. Christensen, R. Liang, D. A. Bonn, W. N. Hardy, U. R\"{u}tt, O. Gutowski, M. v. Zimmermann, E. M. Forgan, and S. M. Hayden, Phys. Rev. Lett. \textbf{110}, 137004 (2013).

\bibitem{Keimer2014} M. Le Tacon, A. Bosak, S. M. Souliou, G. Dellea, T. Loew, R. Heid, K. P. Bohnen, G. Ghiringhelli, M. Krisch, and B. Keimer, Nat. Phys. \textbf{10}, 52 (2014).

\bibitem{Blanco2014} S. Blanco-Canosa, A. Frano, E. Schierle, J. Porras, T. Loew, M. Minola, M. Bluschke, E. Weschke, B. Keimer, and M. Le Tacon, Phys. Rev. B \textbf{90}, 054513 (2014).

\bibitem{Gerber2015} S. Gerber, H. Jang, H. Nojiri, S. Matsuzawa, H. Yasumura, D. A. Bonn, R. Liang, W. N. Hardy, Z. Islam, A. Mehta, S. Song, M. Sikorski, D. Stefanescu, Y. Feng, S. A. Kivelson, T. P. Devereaux, Z.-X. Shen, C.-C. Kao, W.-S. Lee, D. Zhu, and J.-S. Lee, Science \textbf{350}, 949 (2015).

\bibitem{Chang2016} J. Chang, E. Blackburn, O. Ivashko, A. T. Holmes, N. B. Christensen, M. Hucker, R. Liang, D. A. Bonn, W. N. Hardy, U. Rutt, M. v. Zimmermann, E. M. Forgan, and S.M. Hayden, Nat. Commun. \textbf{7}, 11494 (2016). 

\bibitem{Jang2016} H. Jang, W.-S. Lee, H. Nojiri, S. Matsuzawa, H. Yasumura, L. Nie, A. V. Maharaj, S. Gerber, Y. J. Liu, A. Mehta, D. A. Bonn, R. Liang, W. N. Hardy, C. A. Burns, Z. Islam, S. Song, J. Hastings, T. P. Devereaux, Z. X. Shen, S. A. Kivelson, C.-C. Kao, D. Zhu, and J.-S. Lee, Proc. Nat. Acad. Sci. \textbf{113}, 14645 (2016).

\bibitem{Alonso2015} R. Alonso-Mori \textit{et al.}, J. Syn. Rad. \textbf{22}, 508 (2015).

\bibitem{Haceker2013} M. H\"{u}cker, M. v. Zimmermann, Z. J. Xu, J. S. Wen, G. D. Gu, and J. M. Tranquada, Phys. Rev. B \textbf{87}, 014501 (2013).

\bibitem{Croft2014} T. P. Croft, C. Lester, M. S. Senn, A. Bombardi, and S. M. Hayden, Phys. Rev. B \textbf{89}, 224513 (2014).

\bibitem{Comin2014} R. Comin, A. Frano, M. M. Yee, Y. Yoshida, H. Eisaki, E. Schierle, E. Weschke, R. Sutarto, F. He, A. Soumyanarayanan, Y. He, M. Le Tacon, I. S. Elfimov, J. E. Hoffman, G. A. Sawatzky, B. Keimer, and A. Damascelli, Science \textbf{343}, 390 (2014).

\bibitem{Neto2014} E. H. da Silva Neto, R. Comin, F. He, R. Sutarto, Y. Jiang, R. L. Greene, G. A.  Sawatzky, and A. Damascelli, Science \textbf{343}, 393 (2014).

\bibitem{Tabis2014} W. Tabis, Y. Li, M. Le Tacon, L. Braicovich, A. Kreyssig, M. Minola, G. Dellea, E. Weschke, M. J. Veit, M. Ramazanoglu, A. I. Goldman, T. Schmitt, G. Ghiringhelli, N. Bari\v{s}i\'{c}, M. K. Chan, C. J. Dorow, G. Yu, X. Zhao, B. Keimer, and M. Greven, Nat. Commun. \textbf{5}, 5875 (2014).

\bibitem{Neto2015} E. H. da Silva Neto, P. Aynajian, A. Frano, R. Comin, E. Schierle, E. Weschke, A. Gyenis, J. Wen, J. Schneeloch, Z. Xu, S. Ono, G. Gu, M. Le Tacon, and A. Yazdani. Science \textbf{347}, 282 (2015).

\bibitem{Lalibert2017} F. Lalibert\'{e}, M. Frachet, S. Benhabib, B. Borgnic, T. Loew, J. Porras, M. Le Tacon, B. Keimer, S. Wiedmann, Cyril Proust, D. LeBoeuf, https://arxiv.org/abs/1705.07763.

\bibitem{Haug2010} D. Haug, V. Hinkov, Y. Sidis, P. Bourges, N. B. Christensen, A. Ivanov, T. Keller, C. T. Lin, and B. Keimer, New J. Phys. \textbf{12}, 105006 (2010).

\bibitem{Blanco2013} S. Blanco-Canosa \textit{et al.}, Phys. Rev. Lett. \textbf{110}, 187001 (2013).

\bibitem{Nie2017} L. Nie, A. V. Maharaj, E. Fradkin, and S. A. Kivelson, Phys. Rev. B \textbf{96}, 085142 (2017).


\end{thebibliography}
\end{document}